\documentclass[sigplan, screen]{acmart}

\usepackage{algorithm}
\usepackage[noabbrev]{cleveref}
\usepackage{amsmath,amsfonts}
\usepackage{enumitem}
\usepackage{listings}
\usepackage{xspace}

\usepackage{titlesec}

\usepackage{tcolorbox}
\tcbuselibrary{listings,skins}
 
\newcommand{\ie}{\hbox{\emph{i.e.,}}\xspace}

\definecolor{Code}{rgb}{0,0,0} 
\definecolor{Decorators}{rgb}{0.5,0.5,0.5} 
\definecolor{Numbers}{rgb}{0.5,0,0} 
\definecolor{MatchingBrackets}{rgb}{0.25,0.5,0.5} 
\definecolor{Keywords}{rgb}{0.5,0,0.5} 
\definecolor{self}{rgb}{0,0,0} 
\definecolor{Strings}{rgb}{0,0.63,0} 
\definecolor{Comments}{rgb}{0,0.63,1} 
\definecolor{Backquotes}{rgb}{0,0,0} 
\definecolor{Classname}{rgb}{0,0,0} 
\definecolor{FunctionName}{rgb}{0.45,0.30,0.15} 
\definecolor{Operators}{rgb}{0,0,0} 
\definecolor{Background}{rgb}{0.98,0.98,0.98} 
 
\lstdefinelanguage{Python}{ 
    xleftmargin=1em, 
    xrightmargin=1em,
    framextopmargin=0.5em, 
    framexbottommargin=0.5em, 
    showspaces=false, 
    showtabs=false, 
    showstringspaces=false, 
    frame=l, 
    tabsize=4, 
    basicstyle=\ttfamily\small\setstretch{1}, 
    backgroundcolor=\color{Background}, 
    commentstyle=\color{Comments}\slshape, 
    stringstyle=\color{Strings}, 
    morecomment=[s][\color{Strings}]{"""}{"""}, 
    morecomment=[s][\color{Strings}]{'''}{'''}, 
    morekeywords={import,from,class,def,for,while,if,is,in,elif,else,not,and,or,print,break,continue,return,True,False,None,access,as,,del,except,exec,finally,global,import,lambda,pass,print,raise,try,assert,len,tuple, list}, 
    keywordstyle={\color{Keywords}\bfseries}, 
    morekeywords={[2]fuzz_sum}, 
    keywordstyle={[2]\color{FunctionName}\bfseries}, 
    emph={self}, 
    emphstyle={\color{self}\slshape}, 
}

\definecolor{blueish}{RGB}{250, 250, 255}
\definecolor{greenish}{RGB}{250, 255, 250}

\definecolor{gray05}{gray}{0.95}
\definecolor{gray08}{gray}{0.92}
\definecolor{gray10}{gray}{0.90}
\definecolor{gray12}{gray}{0.88}
\definecolor{gray15}{gray}{0.85}
\definecolor{gray18}{gray}{0.82}
\definecolor{gray20}{gray}{0.80}
\definecolor{gray25}{gray}{0.75}
\definecolor{gray30}{gray}{0.70}
\definecolor{gray35}{gray}{0.65}
\definecolor{gray40}{gray}{0.60}
\definecolor{gray45}{gray}{0.55}
\definecolor{gray50}{gray}{0.50}
\definecolor{gray55}{gray}{0.45}
\definecolor{gray60}{gray}{0.40}
\definecolor{gray65}{gray}{0.35}
\definecolor{gray70}{gray}{0.30}
\definecolor{gray75}{gray}{0.25}
\definecolor{gray80}{gray}{0.20}
\definecolor{gray85}{gray}{0.15}
\definecolor{gray90}{gray}{0.10}
\definecolor{gray95}{gray}{0.05}

\newcommand{\fus}{Full-Seq\xspace}
\newcommand{\fos}{First-Of-Seq\xspace}

\newtcbox{\inlinebox}[1][]{enhanced,
 box align=base,
 nobeforeafter,
 colback=blueish,
 size=small,
 left=0pt,
 right=0pt,
 boxsep=2pt,
 #1}
 
\newcommand{\highlight}[1]{%
{\footnotesize%
\inlinebox{#1}%
}}

\usepackage{algorithm}
\usepackage[noend]{algpseudocode}

\makeatletter
\algrenewcommand\ALG@beginalgorithmic{\small}
\makeatother
\algnewcommand{\LineComment}[1]{\State \(\triangleright\) #1}

\title{Predictive Synthesis of API-Centric Code}

\setcopyright{acmcopyright}
\acmPrice{15.00}
\acmDOI{10.1145/3520312.3534866}
\acmYear{2022}
\copyrightyear{2022}
\acmSubmissionID{pldiws22mapsmain-p5-p}
\acmISBN{978-1-4503-9273-0/22/06}
\acmConference[MAPS '22]{Proceedings of the 6th ACM SIGPLAN International Symposium on Machine Programming}{June 13, 2022}{San Diego, CA, USA}
\acmBooktitle{Proceedings of the 6th ACM SIGPLAN International Symposium on Machine Programming (MAPS '22), June 13, 2022, San Diego, CA, USA}

\begin{document}


\author{Daye Nam}
\affiliation{%
  \institution{Carnegie Mellon University$\dagger$\thanks{$\dagger$ Work done at Facebook as an intern}}
  \country{U.S.A.}  
}

\author{Baishakhi Ray}
\affiliation{%
  \institution{Columbia University$\dagger$\thanks{$\dagger$ Work done at Facebook as visiting scientist; equal contribution as the first author}}
  \country{U.S.A.}
}


\author{Seohyun Kim}
\affiliation{%
  \institution{Meta}
  \country{U.S.A.}
}

\author{Xianshan Qu}
\affiliation{%
  \institution{Meta}
  \country{U.S.A.}  
}

\author{Satish Chandra}
\affiliation{%
  \institution{Meta}
  \country{U.S.A.}  
}



\begin{abstract}



Today's programmers, especially data science practitioners, make heavy use of data-processing libraries (APIs) such as PyTorch, Tensorflow, NumPy, and the like.  Program synthesizers can provide significant coding assistance to this community of users; however program synthesis also can be slow due to enormous search spaces.  


In this work, we examine ways in which machine learning can be used to accelerate enumerative program synthesis.  
We present a deep-learning-based model to predict the sequence of API functions that would be needed to go from a given input to a desired output, both being numeric vectors.  Our work is based on two insights. First, it is possible to learn, based on a large number of input-output examples, to predict the likely API function needed.
Second, and importantly, it is also possible to learn to \emph{compose} API functions into a sequence, given an input and the desired final output, without explicitly knowing the intermediate values.

We show that we can speed up an enumerative synthesizer by using predictions from our model variants.   These speedups significantly outperform previous ways (e.g. DeepCoder~\cite{balog2016deepcoder}) in which researchers have used ML models in enumerative synthesis.





\end{abstract}

\maketitle


\section{Introduction}
\label{sec:intro}

One of the cherished dreams of the programming languages research community is to enable automated synthesis of programs based on a  specification.  Synthesis approaches have been designed around several different forms of specification, e.g. a formal specification, or natural language description, or input-output examples (aka demonstration), or a combination thereof.  Just as well, several different approaches to synthesis have been researched; see Related Work.

Our focus is on coding assistance for users of numeric libraries such as PyTorch, Tensorflow, Numpy, Pandas, and the like, each of which provide powerful data manipulation routines behind an API, and the API functions are generally side-effect free.  We assume a specification in the form of a single input-output example, and we are looking for a straight-line program consisting of calls to API functions.  We choose \textit{enumerative synthesis} (explained in the next subsection) as the underlying synthesis approach. Our research goal---shared with recent works such as DeepCoder~\cite{balog2016deepcoder}, TF-Coder~\cite{shi2020tf}, Autopandas~\cite{bavishi2019autopandas}, and others---is to speed up plain enumerative synthesis using machine learning (ML).  


Here is an input matrix as well as the desired output matrix, and the synthesis problem is to come up with a sequence of function calls that would convert the input to output.  We will use the PyTorch API for this purpose.

\begin{small}
\begin{verbatim}
  in = [[5., 2.], [1., 3.], [0., -1.]]
  out = [[[5., 5.], [1., 1.], [0., 0.]],
         [[2., 2.], [3., 3.], [-1., -1.]]]
\end{verbatim}
\end{small}

\noindent
The desired code fragment for this example is: 
\noindent
\begin{small}
\begin{verbatim}
  transpose(stack((in, in), 2), 0, 1)
\end{verbatim}
\end{small}

The goal of program synthesis is to arrive at this expression, given only the input and output.  Keep in mind that
it is unlikely that random guessing of an expression will work: there are tens if not hundreds of available functions, and each function might take more than one argument.  Thus, a systematic search is necessary.

\begin{figure*}
    \centering
    \includegraphics[scale=0.5]{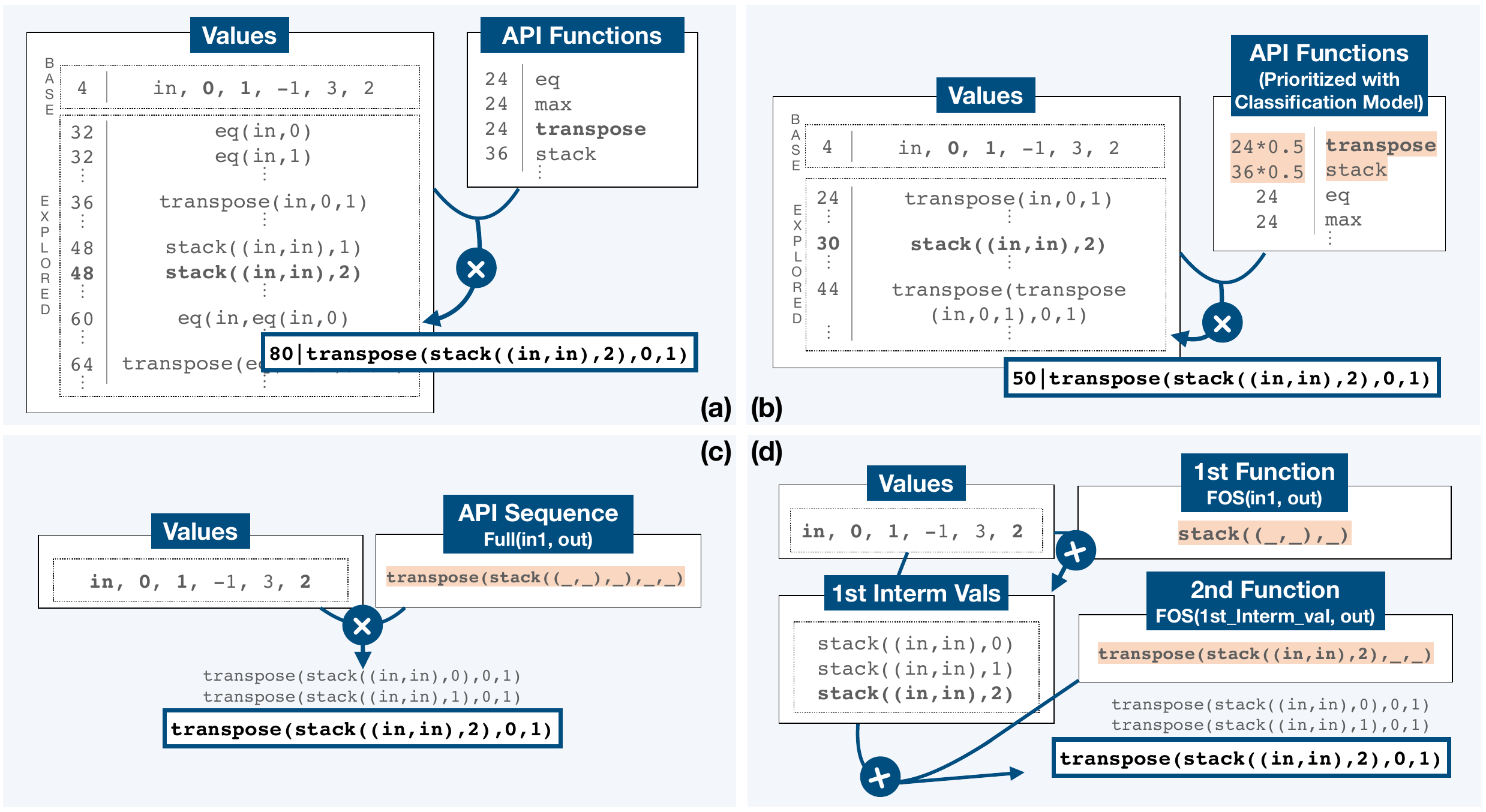}
    \caption{Overview of ML guided enumerative search algorithms. 
    (a) weighted enumerative synthesis without ML model incorporation~\cite{shi2020tf}, 
    (b) weighted enumerative synthesis with one-time ML-based prioritization~\cite{shi2020tf,balog2016deepcoder}, (c) incorporation of \fus prediction mode, (d) incorporation of \fos prediction mode. Red-highlight indicates the API functions predicted by ML models. The underscore is a placeholder for argument values.
    Numbers (in (a),(b)) on the left side are the costs assigned to the values and API functions.}
    \label{fig:overview}
\end{figure*}

\paragraph*{Basic enumerative synthesis} Refer to Figure~\ref{fig:overview}, part (a), where we illustrate an enumerative synthesis in the style of Transit~\cite{udupa2013transit} and TF-Coder~\cite{shi2020tf}. The idea is to organize the
search in the order of increasingly complex expression trees,
where the complexity is approximated by a \textit{cost}.  We assign a cost to each available value, and to each operation, which here are API functions.  (The cost of a function is assigned heuristically, e.g. based on a global frequency of usage.) At each step,
we work with a budget, which grows in successive steps.  Expressions that can be formed from existing values within the budget are added to a pool of values.  For instance, 
the expression {\small{\texttt{stack((in,in),2)}}} would cost the two times the cost of {\small{\texttt{in}}} plus the costs of the value {\small{\texttt{2}}} and the function {\small{\texttt{stack}}}.  In the figure, this cost comes to 48, based on the cost of {\small{\texttt{stack}}} being 36. The value computed by this expression is added to the pool of values, along with the expression that computes them and its cost.  The process continues, with increasing budget at each step, until the desired output value is found.  The expressions added to the pool of values are shown in the figure. 

\paragraph*{Trying likely functions first} The enumerative search presented above is slow, and gets exponentially slower if a larger expression is needed to get the job done.  A reason for this slowness is that the turn of the actually needed API function might come in quite late, as enumerative synthesis makes its way through the smaller cost budget and cheaper functions.  Balog et al~\cite{balog2016deepcoder}, in their seminal work DeepCoder, described a machine learning based strategy to accelerate enumerative program synthesis. 
DeepCoder's insight is to re-assign costs to functions---based on a machine learning model over the given input and output---such that the function(s) more likely to be needed in a \textit{given} situation are prioritized.  See part (b) of Figure~\ref{fig:overview}. Here, given the specific input and output, DeepCoder's machine learning model, adapted to our setting, correctly deems {\small{\texttt{transpose}}},  and {\small{\texttt{stack}}} as likely to be needed.  Operationally, an enumerative synthesis process (e.g. as implemented in TF-Coder~\cite{shi2020tf}) can  lower costs of these operations by some factor, so they are likely to be tried in preference to other API functions.  The hope is that if the ML prediction is accurate, and the discounted costs work out, the process of enumerative synthesis can be sped up considerably.


\paragraph*{This paper: Predicting function sequences} Our thesis is that ML can be used in the setting of enumerative synthesis of API-centric code in a more powerful way: not for prioritization, but instead to directly predict the sequence of API functions that required 
to go from input(s) to the desired output.  We describe two ways in which such a predictive model can be used to accelerate enumerative synthesis.

The first way in which we use this prediction model is to just let it predict the entire sequence of API functions in one shot, given the input and the output. In our running example, the model will predict {\small{\texttt{stack, transpose}}} as the sequence.  See Figure~\ref{fig:overview}, part (c).  Given this sequence, the enumerative synthesizer will only look for  values to fill into the function call arguments (shown by "\_"). If the model predicts correctly, the search space that an enumerative synthesizer faces is vastly reduced, leading to possibly significant speedups.  

A second way in which we use this predictive model is to use it as a ``first-of-sequence'' (FOS) predictor. See Figure~\ref{fig:overview}, part (d).  Given the input and the desired output, the FOS predictor only predicts the \textit{first} function in the sequence needed. Say it predicts that function is {\small{\texttt{stack}}}.  The synthesizer tries out a set of concrete arguments for  {\small{\texttt{stack}}} from the values pool.  The result of evaluating  {\small{\texttt{stack}}} on each of these sets of arguments is added to the values pool; these are intermediate values in the desired computation.  Next, for each intermediate value thus obtained, the synthesizer invokes the model again, this time giving it the intermediate value (in place of the input) and the desired output value.  Say the model now predicts that the first function in the \textit{remaining} sequence needed is {\small{\texttt{transpose}}}.  The synthesizer then looks for appropriate arguments for {\small{\texttt{transpose}}}.  At this point, one of the argument choices would provide the desired output.  Compared to the full prediction, the point of this FOS mode is that it gets to predict on the basis on \textit{known} intermediate values, a bit akin to teacher forcing~\cite{williams1989learning} in sequence prediction, and can be successful more often than the full prediction mode; but it can be less efficient than one-shot prediction of the entire sequence.

On the running example, here are the comparative times to a successful solution: plain enumerative synthesis, 54.79 seconds; DeepCoder-style ML-based prioritization, 34.71 seconds; our API sequence prediction, FOS mode, 0.49 seconds; and API sequence prediction, Full mode, 1.45 seconds.
In this example, full sequence prediction mode took a tad longer than the FOS mode: this is because the correct full sequence was in top-3 but not top-1, whereas in the FOS model the correct choice were at top-1.  
In general, we have found the full sequence mode to be faster than FOS mode.
Other examples of \fus guided synthesis are available in Table~\ref{tab:prediction}.

\begin{table}
\caption{Sample of synthesized programs with \fus model guided enumerative synthesis and the synthesis time comparison. More examples can be found in the supplementary material (Appendix B.1).}
\label{tab:prediction}
\centering
\footnotesize
\resizebox{\columnwidth}{!}{
\begin{tabular}{c|c||c}
\toprule
 Synthesized Program & \fus (s) & no ML (s)  \\
\hline
\path{eq(in1,unsqueeze(in1,1))}  & 0.18& 0.8 \\ 
\path{tensordot(in1,transpose(in2,0,1),1)} & 0.32& 2.06  \\ 
\bottomrule
\end{tabular}
}
\end{table}





\paragraph*{Contributions}

We make two contributions in this work.
First, we present a way to incorporate powerful  predictive models in the context of enumerative program synthesis.  On a suite of benchmarks (adapted from Stack Overflow) for PyTorch, using our ML models reduces the (mean, max) synthesis time from (10.01, 96.53) to (1.04, 9.58). By contrast, an adaptation of the idea of DeepCoder~\cite{balog2016deepcoder} reduces the (mean,max) synthesis time only to (7.44, 77.00).  (See Section~\ref{subsec:eval_synthesis}, Table~\ref{tab:synthesis}.)
    

Second, our main technical advancement is in being able to carry out prediction of a sequence of API functions, given the input and the final desired output.
Specifically, our model predicts one API function at a time and executes each predicted API function to convert the (intermediate) input state into another intermediate state until it becomes the target output state.
Here, the intermediate states are not given to the model, but the model learns to represent what \emph{would be} concrete intermediate values in the latent space during the training time.
The ability to execute the API functions in the latent space indicates that the model learns the API function semantics (i.e., the relation between the input and output states) rather than the sequence distribution of the training dataset, and allows the model to generalize to unseen sequences or lengths.
See Sec~\ref{sec:whycomp}.



\section{Learning to predict API sequences}
\label{sec:intuition}


Our technique works based on supervised
learning over a large number of input and output examples, trained over individual API functions, or on sequences of API functions.   Since the availability of real training data is a pervasive problem in ML,  we use synthetic data generation similar to prior program synthesis work~\cite{shin2019synthetic}.
We pipe randomly-generated diverse inputs through sequences of API functions and collect resulting outputs (see Section~\ref{sec:fuzzing}).  This helps capture the behavior of a single or a sequence of API functions in terms of how it transforms its input to the output.


Once trained, the model is able to predict a sequence of API functions.  It can predict for input-output pairs that were never seen in the training data; thus it generalizes in the data space as long as the query input-output pairs are in distribution.   More interestingly, it can predict sequences of API functions that were not seen in the training set either.  This latter point is crucial, because the way we train the model, it learns to \emph{compose} new, previously unseen sequences from the behaviors learned from training sequences.

Before we present operational details (Sec~\ref{sec:method} onwards), we would like to present some intuitions behind our proposed ideas.
We start with a basic classification model designed to predict one API function, given an input and desired output; and then build over it a compositional model that is designed to predict a sequence of API functions. 
The importance of examining the classification model on its own was crucial in our own journey, because it helped overcome several challenges in synthetic data generation for training. 
(In actual synthesis application, we use the model that predicts function sequences, described after this.)

\paragraph*{Predicting a function from input-output data}

The first intuition we use is that for many common API functions, their behavior---the relationship of output to inputs---has simple patterns.  Moreover, the behavior of a function is \textit{discriminable} from behaviors of other functions based on simple clues.  Many functions simply move around elements of a data structure (e.g. {\small \texttt{transpose}} or {\small \texttt{reverse}}) in easy to recognize patterns.  In other cases, the operation is a simple element-wise computation.\footnote{There do exist operations with more complex behaviors, but here we limit ourselves to simple ones (see Appendix A.}  This suggests that a feed-forward neural network can be trained to predict likely functions---as in a multi-classification problem---from a representation of the input and output data.  Such a network would have to be trained on large amounts of input-output examples and their known (ground truth) functions.

\begin{figure}
\includegraphics[width=0.8\columnwidth]{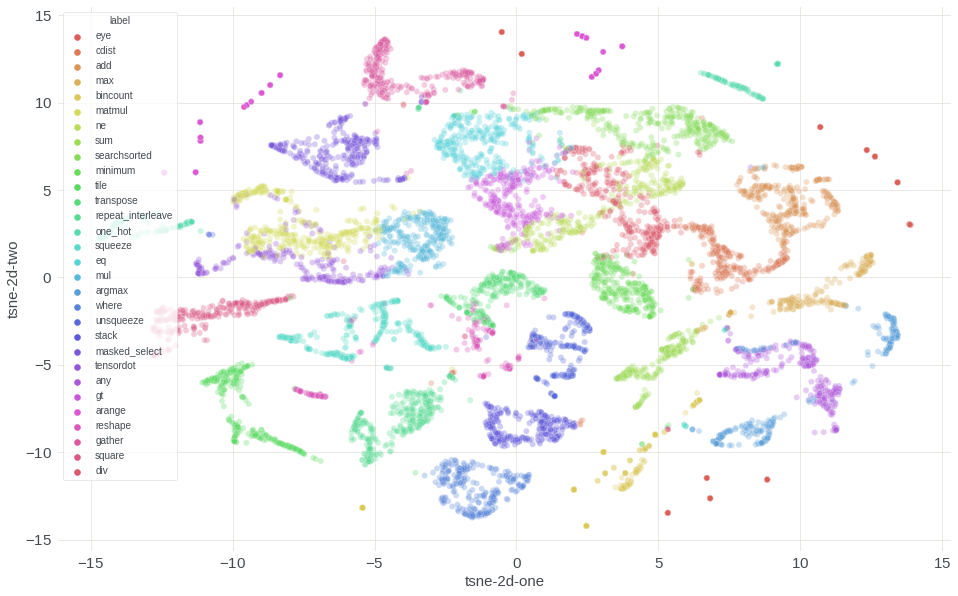}
\caption{Visualization of embedding space of input-output pairs.}\label{fig:tsne-single}
\end{figure}

Figure~\ref{fig:tsne-single} shows the tSNE plot for 5000 input-output pairs. The classification model was trained over synthetic data generated to classify among one of 33 API calls from PyTorch.  The figure shows that the input-output pairs -- or rather, their embeddings -- map to visually distinct clusters, corresponding to the function calls that would be needed to go from the input to the output.  The reason this clusters cleanly is that the network learn to pick up the essential patterns that appear in the input and corresponding outputs.

\paragraph*{Predicting sequences of functions}

The case of predicting an API \textit{sequence}, such as \texttt{stack}
followed by \texttt{transpose}, is harder.  
The intermediate values that flow between API calls are not known ahead of time, so it is not possible to reconstitute this sequence simply by invoking the classification model (for one API method name) over successive pairs of inputs and outputs.  
Moreover, learning to recognize the intended sequence from among all possible sequences, based on an input and the final output, can be difficult, for reasons for computational cost, for a classification model that predicts over a fixed collection of sequences of API function names. 


This is where a second intuition comes into play.   Given an input and a final output, we can imagine a model that predicts the \textit{first} function in the intended sequence of the API functions that would process the input and eventually produce the (final) output. Crucially, we train this model as a \textit{recurrent} unit, such that it not only predicts the first API function needed, but additionally produces a representation (in the embedding space)  of the output of that first API function. (The correspondence of the internal representations to intermediate values is further explored in Section~\ref{sec:whycomp}.)  This representation, along with the final output, can then be passed to a \textit{recurrent} invocation of the model, to make it predict the next API function in the sequence.  In this way, we can train a compositional model for API sequences.


In our running example, the model first predicts {\small \texttt{stack}} based on
{\small{\texttt{in}}} and {\small{\texttt{out}}}.  Importantly, it also computes an internal representation of the intermediate value {\small \texttt{stack((in,in),2)}}.   It then predicts {\small \texttt{transpose}} based on this internal representation and {\small \texttt{out}}. This is the principle by which the model is able to compose even longer previously-unseen sequences. 


Here we emphasize that the model is \emph{not} predicting the next API token (e.g. {\small \texttt{transpose}}) based on the tokens that came before (e.g. {\small \texttt{stack}}), as is done in code completion models~\cite{hindle}.  At each step, the prediction is based \emph{only} on (an internal representation of) the program state, as opposed to on program text.  This is a new capability,  which could optionally be combined with additional signals such as previous tokens, if desired.

\section{Technical Details}
\label{sec:method}
In this section, we explain our models, the training and the inference.
We will use Figure~\ref{fig:example} to show details using an example.

\subsection{Notations}
\label{sec:classification}


We will work with following entities:

\begin{itemize}[leftmargin=*]
    \item $\mathbb{T}$ for domain values, 
    tensors or vectors (or lists thereof)
    \item $\mathbb{E}$ for embeddings, which are vector representations internal to a neural network
    \item $\mathbb{D}$ for distributions, which are probability distributions over names of API functions. For $d \in \mathbb{D}$, $d(f)$ is the probability of function $f$.  
\end{itemize}
We will use some auxiliary operators:
\begin{itemize}[leftmargin=*]
    \item embedding, denoted by $\llbracket . \rrbracket: \mathit{encoding}(\mathbb{T}) \rightarrow \mathbb{E}$
    \item concatenation, denoted by $.\#.: \mathbb{T} \times \mathbb{T} \rightarrow \mathbb{T}$
\end{itemize}

\subsection{The \textit{encoding} function}
\label{sec: encoding}

Before passing the input and output tensors to the models, we encode them into a fixed-length vector (Figure~\ref{fig:example}-Encoding).  We extract three different pieces of information from the tensors: (i) tensor values, (ii) tensor shapes, and (iii) tensor types, and combine them 
as a sequence separated by a special separator $<s>$, \ie  \highlight{$X = $ type~{\tt <s> }~shape~{\tt <s>}~value}, such that, the models can learn from all the three modalities together. 



To manage the wide range of tensor values in the model, we normalize the values as follows: we encoded the values greater than 100 into 100, values greater than 1000 into 101, and similarly for the negative values.
The intuition is based on how developers recognize patterns: when a value becomes large enough, the importance of the least significant digit decreases in pattern recognition.


Finally, all domain inputs and output encoding are concatenated together. 
We support up to 3 inputs and one output. 
Dummy inputs are added when there are less then 3 inputs to keep the model input size same for all examples.





\subsection{Compositional Model}
\label{sec:compositional}

\begin{figure*}[t]
    \centering
    \includegraphics[width=\textwidth]{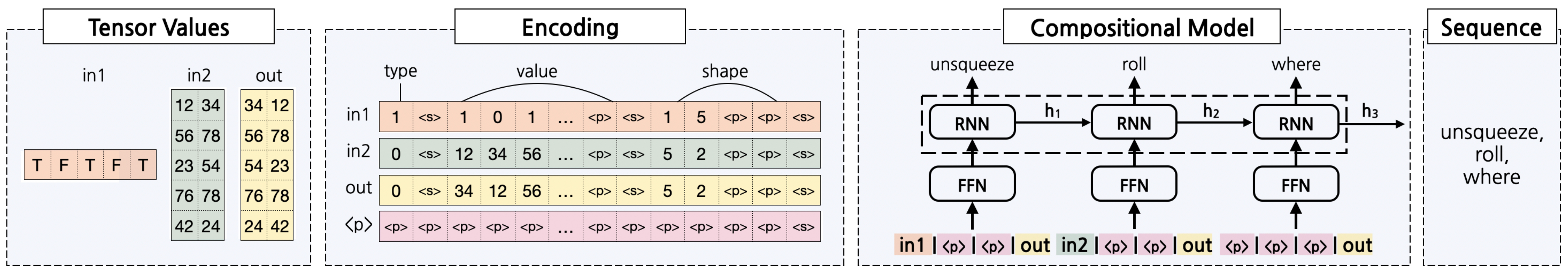}
    \caption{Illustration of Compositional Model on an example. The inputs are in the Tensor Values box, and the expected prediction is shown in the Sequence box.}
    \label{fig:example}
\end{figure*}

We train a model to predict the sequence of API functions $s_f = [f_1, ..., f_n]$, given a task specification $\phi = \{inp, out\}$, where $inp$ is a list of input tensors that have gone through the sequence of API operations $s_f$, and $out$ is the final output tensor.

In this description, we assume all functions take two inputs: first, the result of previous computation, and second, a ``local'' input, e.g. $inp_i$ here that comes from $inp$.
Define $args_i = (f_{i-1}(args_{i-1}), inp_i)$, for $i>1$, and $args_1 = (\_, inp_1)$.
We train a model $G$, such that for $i=1..n$:
\begin{equation}
  f_i = G(\llbracket args_i, out  \rrbracket)
\end{equation}






The embedding $\llbracket . \rrbracket$ of the encoded inputs is obtained using feed-forward networks (FFN), whereas $G$ is rendered 
 by employing recurrent neural networks (RNNs)\footnote{Technically, bi-directional RNNs~\cite{schuster1997bidirectional}.}.  The operation of \emph{one} cell of an RNN has the type:
\(
  \mathit{RNN}:  \mathbb{E}_1 \times \mathbb{E}_2 \rightarrow (\mathbb{E}_3, \mathbb{D})  
\)
where,
\begin{itemize}[leftmargin=*]
    \item $\mathbb{E}_1$ is the hidden state coming from previous cell, or a zero value;
    \item $\mathbb{E}_2$ is the embedding of the local input and the final output; we permit each function to have an optional additional input;
    \item $\mathbb{E}_3$ is the output hidden state being passed to the next cell;
    \item $\mathbb{D}$ is the prediction of API function from this cell; technically it is a distribution from which we take the argmax.
\end{itemize}
Then,
\begin{equation}
    \mathbf{h}_i, d_i = \mathit{RNN}_i(\mathbf{h}_{i-1}, \llbracket inp_i \# out \rrbracket)
\end{equation}

where $\mathbf{h_{i-1}}$ and $\mathbf{h_{i}}$ are incoming and outgoing hidden states, respectively, and $d_i$ the predicted distribution.  We expect $f_i = \mathit{argmax}(d_i)$.

Figure~\ref{fig:example} Compositional Model shows three units of the model for an example.
In each unit, the encoding passed to the feed forward network is similar to the one used before to create $\llbracket inp_{i}\#\mathit{out} \rrbracket$.
When $f_i$ needs to use $f_{i-1}(args_{i-1})$, we mask the position as empty ("<p>" in the figure) so that the model exploits $\mathbf{h}_{i-1}$.   Embedded encodings are passed to RNN units, and each unit further projects the input embedding into the RNN embedding space to generate $h_i$, using information flowed from adjacent units, $h_{i-1}$.
Finally, the output of each unit is passed to a softmax layer (not shown here) 
to produce a probability distribution over API functions.


\subsection{Synthetic Data Generation}
\label{sec:fuzzing}

To train a neural model so that it can understand the behavior of API functions, a large number of corresponding input-output pairs is necessary.
Unlike other problems exploiting ML models, collecting real-world data from code repositories (e.g., GitHub) is not applicable here because we need runtime values, not static information such as static code. Therefore, we randomly generate input/output values, and use the synthetic dataset for model training.

For each API function, we randomly generate input tensors, run the API functions with them, and capture the corresponding outputs.
In other words, we create a set of input/output values in a black-box manner: we do not assume API functions' implementation details or internal behaviors.
As it does not require understanding internal program structures, it is easy to generate a large number of input-output pairs without much manual effort and can be easily parallelized.

\lstset{language=Python}
\lstset{frame=lines}
\lstset{caption={Example data generation code for \texttt{torch.sum}}}
\lstset{label={code:fuzzing}}
\lstset{basicstyle=\footnotesize}
\begin{lstlisting}
def generate_sum_IO():
    in_tensor, tensor_size = random_tensor()
    dim = random_dimension(0, tensor_size)
    if dim == len(tensor_size):
        out_tensor = torch.sum(in_tensor)
    else:
        out_tensor = torch.sum(in_tensor, dim)
    return (in_tensor, out_tensor)

\end{lstlisting}

However, as even a simple API operation in modern libraries (e.g., PyTorch) imposes many constraints, inputting random values will generate many runtime errors due to the constraints violations. 
To reduce such errors, 
we exploit API specification, and generate a set of inputs with the valid combinations (see Listing~\ref{code:fuzzing} for an example of generating datapoints for \texttt{torch.sum}). 
By excluding invalid combinations of arguments to each API function, we can speed up the data generation and generate a large synthetic dataset that can capture API function input/output benign behavior. 

\section{Incorporating ML in Enumerative Syn.}
\label{sec:synthesis}
Here 
we formally describe how the ML models were incorporated into the enumerative synthesis.  Please refer to Figure~\ref{fig:overview} and Section~\ref{sec:intro} for a walk through of these on an example.
Detailed description of our implementation and the pseudo code for each synthesis approach can be found in the supplementary material (Appendix C, D).

\textbf{Basic enumerative synthesis.} 
As a baseline, we implement an enumerative synthesizer without 
any ML models.
Basic enumerative search starts with a set of base values and enumerates over combinations of operations and the values.

The list of base values includes {\small \texttt{inp}}, other basic constants such as {\small \texttt{0}, \texttt{1}, \texttt{-1}}, or heuristically-chosen values such as the dimensions of the given variables (e.g., {\small \texttt{3}}).
Then, starting with the base values, the search enumerates ways of applying operations to previously-explored values and expand the set of known values.
There are various ways of iterating the operations and the values (e.g. based on syntactic size as in Transit~\cite{udupa2013transit}), but we use weighted enumerative search, which is the approach of  and TF-Coder~\cite{shi2020tf}.
It does so in the order of increasing \emph{cost}.
Operations and values are assigned costs based on their complexity: less common and more complex operations are assigned higher cost, and the common and simple operations are assigned lower cost.
Costs are additive, so common operations and simpler expressions are explored earlier.
The costs are manually set by the synthesizer developers,
but only needed to be set once, and it will be used for all tasks.








\textbf{Prioritizing likely functions with an ML model.}
As the needed operations for a specific problem are not known to the synthesizer ahead of time, the costs seeded in it will not always be ideally suited for all problems.
TF-Coder~\cite{shi2020tf} and DeepCoder~\cite{balog2016deepcoder} address this problem using an ML model to re-weigh all operations before the enumerative search starts.
Given a task specification (i.e., input/output examples), it invokes a multi-label classification model to predict the probability of each needed operation and re-weighs them 
accordingly
with the goal of encountering the needed operations earlier in the search. 
We trained multi-label classification model following DeepCoder~\cite{balog2016deepcoder}.






\textbf{Compositional Model - Full-Sequence.} 
In this mode, compositional model predicts a sequence of API functions $s_f = [f_1, f_2, ..., f_n]$ given the final output $out$, and the inputs to each API function $[inp_1, inp_2, .., inp_n]$\footnotemark.
The synthesizer invokes the compositional model with the specification, predicts a sequence
of operations, and searches only the parameter values (e.g., \textit{dimension}) that were not provided in the specification.

The \fus mode completely bypasses enumerative search over operations.
Instead, the compositional model predicts the API functions needed in a synthesis instance as well as the order of those APIs in the synthesized code.
Thus, the synthesizer does not need to search the operation space, but only needs to search the combinations of base values.

\footnotetext{As the program synthesis task specification only provides an order agnostic list of inputs, the synthesizer needs to search through different combinations of them to generate \textit{a list of input tensors to each API call} to invoke the compositional model. In this section, we assume that the list of input tensors to each API call is provided.}
\textbf{Compositional Model - First-Of-Sequence.}
In the \fos mode, given an input and a final output, compositional model predicts the most probable API function needs to come in the sequence.
As enumerative search keeps track of the intermediate output value, we can iteratively invoke compositional model, and compute the intermediate values using the predicted API functions, which can be used to predict the next API function.


\section{Evaluation}
\label{sec:result}

In this section, we first describe the training and evaluation dataset (~\Cref{subsec:dataset}), and evaluate the trained API function sequence prediction model (~\Cref{subsec:multi_api_1}).
Then, we investigate the prediction-guided synthesis (~\Cref{subsec:eval_synthesis}).
Finally, we show the generalizability and the compositional property of our model (~\Cref{subsec:multi_api_2}).

\subsection{Dataset}
\label{subsec:dataset}
\noindent
\textbf{Program synthesis benchmarks.} 
We evaluated the effectiveness of our approaches with a subset of TF-Coder's SO benchmarks~\cite{shi2020tf}.
These benchmarks contain 50 tensor manipulation examples collected from SO, 
each containing input and output tensor values and the desired solutions in Tensorflow. 
To evaluate our approach that supports PyTorch, we first translated them into PyTorch and excluded tasks that we could not translate by hand.
Among the 33 API functions needed to for remaining 36 benchmarks, we selected 16 functions covering 18 benchmarks (Table~\ref{tab:dataset}-Stack Overflow) from core utility that modify values (e.g., {\small \texttt{add}}) or shapes (e.g., {\small \texttt{transpose}}) of tensors, create them, or manipulate them in similar ways.
These operations were chosen because the model can clearly observe the behavior of each API function solely from input and output pairs (i.e., no side effects). 
The full benchmarks we support are available in the supplementary material (Appendix B).

\noindent
\textbf{Synthetic data generation.} 
To train our sequence prediction model that work for 
the SO benchmarks, we synthesized a dataset as per~\Cref{sec:fuzzing}.
We synthesized 202 unique API functions sequences by using the exhaustive combination of 16 API functions, with 1 or 2-length sequences.
From the 272 (16 + 16*16) possible sequences, 70 sequences were removed due to the constraints.

For each API function sequence in the training dataset, say $f_1, f_2, f_3$, we ran $f_1$ with randomly generated input and other parameter values (e.g., dimension, mode, etc.). 
Then, 
$f_2$ takes $f_1$'s output as input and takes other random input tensors, if necessary. We treat $f_3$ similarly by propagating $f_2$'s output. 

To generate diverse and unbiased input-output pairs, we cover different properties of the functions, such that the model can explore the broad data space of input-output pairs. 



It took 1-person week to encode the API specifications to write valid data generation code by reading the PyTorch documentation. 
To avoid expansion to large input values and to let the model learn the patterns sufficiently, 
we used a fixed range of values (from 0 to 20) and the size of tensors (up to 3 dimensions, and up to 5 elements in each dimension), to prevent the tensors to be dispersed too much.

We created the dataset with 100,000 input-output pairs for each unique API sequence (Table~\ref{tab:dataset}-Synthetic).
We split the dataset into training, validation, and test sets.
The training, validation, and test sets included all 202 API sequences, but the input/output values were not overlapped across the datasets.


\begin{table}
\caption{Statistics of the dataset used in this study. Numbers in parentheses indicate the length of the sequences. }
\label{tab:dataset}
\centering
\footnotesize
\resizebox{\columnwidth}{!}{
\begin{tabular}{c|ccc|c}
\toprule
                               & \multicolumn{3}{c|}{Synthetic}   & Stack Overflow              \\ 
\cmidrule(lr){2-5}
                               & \multicolumn{1}{c}{Train} & \multicolumn{1}{c}{Valid.} & \multicolumn{1}{c|}{Test} & Test \\ 
\midrule
\multicolumn{1}{l|}{\# of unique seqs (len)}   &   & 16 (1) + 186 (2) &  &  8 (1) + 7 (2) \\
\multicolumn{1}{l|}{\# of in/out values}      & 5.5M & 10K & 10K   & 18 \\
\bottomrule
\end{tabular}
}
\end{table}

\subsection{Sequence Prediction Model}
\label{subsec:multi_api_1}
We trained both \fus and \fos variants using the training set of the synthetic data, and evaluated it with (1) the test set of the synthetic data, and (2) SO benchmarks.





\begin{table}
   \caption{Model accuracy for unseen input/output values.}
   \label{tab:rq2acc}
   \footnotesize
   \centering
   \begin{tabular}{lc|cr}
   \toprule
                     & \multicolumn{1}{c|}{\textbf{Synthetic-Test}} & \multicolumn{2}{c}{\textbf{Stack Overflow}} \\
    \textbf{Model}   & \textbf{Top-1}  & \textbf{Top-1} & \textbf{Top-3}  \\ 
   \midrule
   \fus & 79.36\%  & 35.29\% & 76.47\%  \\
   \fos     & 66.88\%      & 52.38\% & 76.19\%   \\
   \bottomrule
   \end{tabular}
\end{table}

\noindent
\textit{Observation.} ~\Cref{tab:rq2acc} shows the result.  
Model's top-1 testing accuracies of the 10K synthetic test set are $\sim$79\%. 
Among 18 SO benchmarks, the \fus model found 13 sequences are in top-3 (72.22\%), among them 6 are in top-1 (33.33\%). 
In comparison to the \fus model, the \fos model's top-1 accuracy is better. 
This is not surprising as \fos model has more information (actual values of the intermediate inputs) than the \fus variant.
However, surprisingly, the top-3 accuracies of both are almost similar. 
These results indicate that the compositional model perhaps learned a representation of the intermediate states of the API operation sequence: even without passing the true intermediate values, the \fus model behaves at per with the \fus model at top-3.

\begin{table}
\caption{End-to-end program synthesis results; our models in \textbf{bold}. Time, Max, and Median show the average, max, median synthesis time of found programs.}
\label{tab:synthesis}
\centering
\footnotesize
\begin{tabular}{l|cc|ccc}
\toprule
&  &  & \multicolumn{3}{c}{Time} \\ \cmidrule{4-6}
& Found & Not Found & Mean & Max & Median \\
\midrule
Enumerative      & 18    & 0    & 10.01 & 96.53 & 0.46 \\
Multi-label      & 18    & 0    & 7.44  & 77.00 & 0.32  \\
\textbf{\fos}             & 17    & 1    & 5.87 & 59.93 & 0.39 \\
\textbf{\fus}     & 14    & 4   & 1.04 & 9.58   & 0.25\\
\bottomrule
\end{tabular}

\end{table}

\subsection{Prediction-guided enumerative synthesis}
\label{subsec:eval_synthesis}
Using the trained \fus and \fos variants, we first evaluate our approach, against vanilla enumerative synthesis (similar to ~\cite{shi2020tf}). 
We further compared with a multi-label prediction model, a setting inspired by DeepCoder.~\cite{balog2016deepcoder}.
Table~\ref{tab:synthesis} shows the results.


\textbf{Existing synthesizers vs. compositional model.} 
Among the 18 tasks, both vanilla enumerative search and a synthesizer prioritized with multi-label classification model could synthesize all tasks. 
The new variants incorporating our models synthesized 17 (\fos) and 14 (\fus) correctly, out of 18 and 17 respectively.
However, although they synthesized fewer solutions, they required less time to synthesize the solutions: 5.87 seconds (\fos) and 1.04 seconds (\fus) on average, whereas the existing synthesizers took 10.01 and 7.44 respectively.

We see this speed up because predicting the sequence reduces the search space. 
As the compositional models return a sequence of API functions, the enumerative search can focus on the argument values instead of iterating over the API function sequences. 
Note that the multi-label classification model also suggests potential API functions, but instead of function sequences, it provides us with a set of functions.  Thus, in the worst case, the associated enumerative search has to explore all the possible combinations increasing the synthesis time.  

The difference in synthesis time between the compositional models and the baselines is not big in simple tasks (e.g., {\small \texttt{any(in,-1)}}), which is why the difference in median in Table~\ref{tab:synthesis} is not significant enough.
However, when it comes to more complex tasks like in Figure~\ref{fig:overview}, the difference becomes significant: 54.79 seconds with plain enumerative synthesis vs. 0.49 with \fos mode.

One caveat of compositional models is that the model prediction is not the bottom-up method but a one-shot approach.
Therefore, when the model fails to predict the sequence correctly, it cannot synthesize the program.
However, as the whole search can be done quickly, the time overhead is not high even when one tries with the compositional model and employ other approaches once it fails.

\textbf{\fos vs.~ \fus.}
~Between the two 
variants, \fos 
was able to synthesize more programs.
For example, 
\fos 
successfully synthesized the desired program 
{\small \texttt{where(lt}} {\small \texttt{(in,1),in,1)}} by predicting {\small \texttt{lt}} and {\small \texttt{where}} correctly in top-3.
However, \fus 
failed, and predicted {\small \path{[[eq,where],[eq,mul],[gt,where]]}} 
as top-3, 
which are close enough, but not entirely correct. 
This is expected. \fos only has to get the first element of the sequence right; 
the next element is in fact the first element of the result of the \emph{subsequent} prediction, which in turn, is based on the actual intermediate value computed by the first function predicted.  
Whereas, the \fus gets only one chance to get the entire sequence right without knowing the intermediate values. 
When a sequence is correctly predicted, \fus model could synthesize the solutions faster; it only needs to be invoked once, without the intermediate values computation.






\subsection{Evaluating Generalization} 
\label{subsec:multi_api_2}


To see whether the model truly learned the functionality of the API functions and learned their compositions, we tested whether the model can generate new API sequences that were not present in the training data.
From the original synthetic dataset~\ref{tab:dataset}, we removed the data of 7 API function sequences with length-2 that were included in Stack Overflow benchmarks, and trained the \fos and \fus models.
Our hypothesis was that if the models are able to learn the compositional property, instead of learning the distributions of sequences, they should be able to generate unseen sequences by composing API functions into a sequence.

                     
 
   

\noindent
\textit{Observation.} 
Not surprisingly, the accuracy drops from the Section~\ref{subsec:multi_api_1} result.
Nevertheless, out of 8 benchmarks with 2 sequence, we can still predict 4 sequences at top-5 (50\% accuracy) with \fus, and 6 sequences (75\% accuracy) with \fos.
In this setting, we sometimes narrowly miss some function sequences. For example, we miss a benchmark [{\small \texttt{lt}}, {\small \texttt{where}}], however it predicts [{\small \texttt{eq}}, {\small \texttt{where}}], and [{\small \texttt{gt}}, {\small \texttt{where}}] instead. Note that {\small \texttt{gt}} and {\small \texttt{eq}} have very similar functionalities to the intended API function {\small \texttt{lt}}. 
Both the models can correctly predict sequences like [{\small \tt unsqueeze}, {\small \tt eq}],  [{\small \tt matmul}, {\small \tt add}], etc. As expected, the \fos model works much better than \fus model.

To further check the model's ability to generalize to unseen 3-length sequence, we randomly picked 71 unique 3-length sequences made out of 16 API functions and collected 100 instances of them with different input/output values. 
This gives a total of 7100 test samples.  
We used the model trained with only sequences with length 2.
Overall, at top-5, model's accuracy is $\sim$34\% when queried with unknown sequences and unknown values. 
However,  the model can predict 69 out of 71 sequences correctly at least with one input-output.
The only two sequences the model missed are [{\small \tt  add}, {\small \tt mul}, {\small \tt any}] and [{\small \tt add}, {\small \tt unsqueeze}, {\small \tt ne}]. 
In contrast, [{\small \tt  where}, {\small \tt  expand}, {\small \tt  matmul}] was predicted correctly around 97\% time. These results indicate the model's ability to generalize. 


\begin{table}
  \caption{Model accuracy for unseen input/output values, trained with a dataset covering all SO benchmarks.}
  \label{tab:scalability}
  \footnotesize
  \centering
  \begin{tabular}{lc|cr}
  \toprule
                     & \multicolumn{1}{c|}{\textbf{Synthetic-Test}} & \multicolumn{2}{c}{\textbf{Stack Overflow}} \\
    \textbf{Model}   & \textbf{Top-1}  & \textbf{Top-1} & \textbf{Top-3}  \\ 
  \midrule
  \fus & 88.15\%  & 68.57\% & 91.42\%  \\
  \fos    & 65.44\% & 51.61\% & 79.03\%   \\   
  \bottomrule
  \end{tabular}
\end{table}

\subsection{Scaling to a larger set of API functions}
\label{subsec:eval_scalability}

\textbf{\fos vs.~ \fus.} We created a dataset covering all 36 SO benchmarks (1- to 3-length) in section~\ref{subsec:dataset} with 33 API functions (up from 16 before).
Synthesizing a training data covering all exhaustive combinations of the 33 API functions, up to 3-length sequences, gives us a huge number of combinations (33x33x33=35,937).  Generating the corresponding training data (>35B samples) and train the model accordingly is challenging.  
Therefore, as a start, we selected 65 unique sequences that can be a solution to one of the 36 benchmarks.

\textit{Observation.} ~\Cref{tab:scalability} shows the result. Compared to the model trained with the exhaustive combinations of 16 API functions (Section~\ref{tab:rq2acc}), the \fus model trained with this dataset actually performed even better for (91.42\% top-3 accuracy in  SO benchmark data).
We conjecture that this is because the model could focus its learning on the semantics of API functions sequences that are more likely to be in the test set, rather than learning semantics of all combinations.
Unlike the accuracy difference in \fus models, \fos variant achieved slightly lower top-1 accuracy with this dataset.
This is because the number of API functions it needs to learn has increased from 16 to 33.
For \fos, as it predicts each API function independent from others, it could not benefit from having a training set with the targeted sequences.

\textbf{Generalizability.}
We also tested the generalizability of our model when it is trained with a dataset covering all 33 API functions needed to synthesize the solutions for 36 SO benchmarks.
Due to the aforementioned challenge, we could not synthesize a dataset like in \Cref{subsec:multi_api_2}; instead, we synthesized a dataset with 598 random combinations of 1 or 2-length sequences of 33 API functions, that has not appeared in SO benchmark.

\textit{Observation.} 7 out of 18 (39\%) two-length unseen sequences were predicted correctly with \fus, and the first API of a sequence correctly predicted by \fos with 69\% accuracy.
They both achieved lower accuracy than the ones trained with the exhaustive combinations, as this model had fewer sequences to learn from, 
which is critical in generating unseen sequences.

\section{Why Composition works?}
\label{sec:whycomp}

We show that a unit of our compositional model has an interesting property: it learns to convert its incoming hidden vector to its outgoing hidden vector in a way consistent with the \textit{semantics} of the API function it predicts, albeit in embedding space.  This property is crucial for predicting a sequence compositionally.

\begin{figure}
\includegraphics[width=0.75\columnwidth]{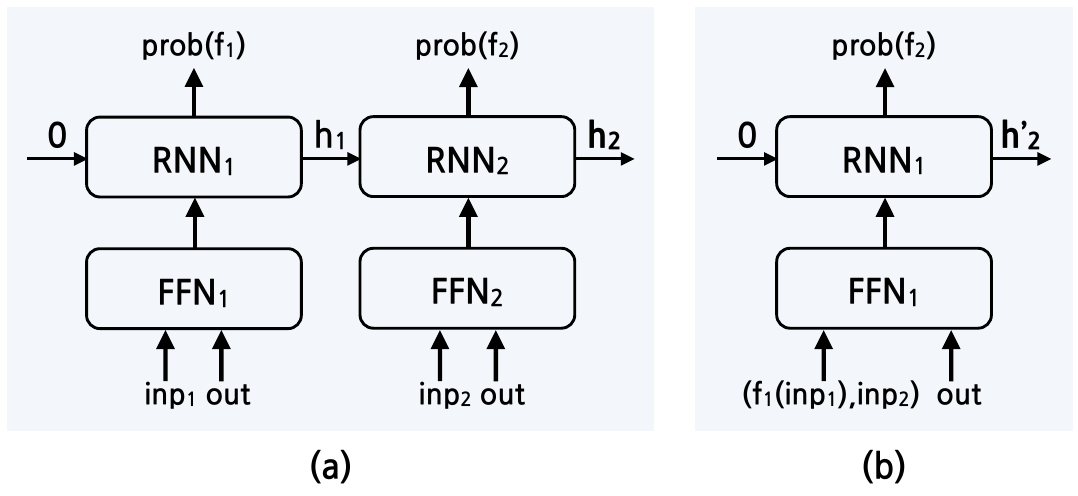}
\caption{
Illustration of compositional learning.  
(a): Two units of the compositional model predicting a sequence [$f_1, f_2$]. 
(b): Single unit model predicting $f_2$ given $f_1(inp_1)$ instead of $h_1$.
We show the compositional property of the model by showing $h_2 \approx h'_2$.
}\label{fig:equivalence2}
\end{figure}

In Fig~\ref{fig:equivalence2}(a), we show two units of the compositional model, where the first one
predicts function $f_1$, on the basis of $inp_1$, $out$ and 
the previous hidden vector, if any.   
That unit also produces a hidden vector $h_1$.  The second unit produces hidden vector $h_2$.  It may also consume further local input (such as $inp_2$).
Fig~\ref{fig:equivalence2}(b) shows an alternate situation in which we give the \emph{result} of $f_1(inp_1)$ directly as input to the first unit, which then produces $h_2'$.

\begin{figure}[!h]
    \centering
    \includegraphics[width=0.75\columnwidth]{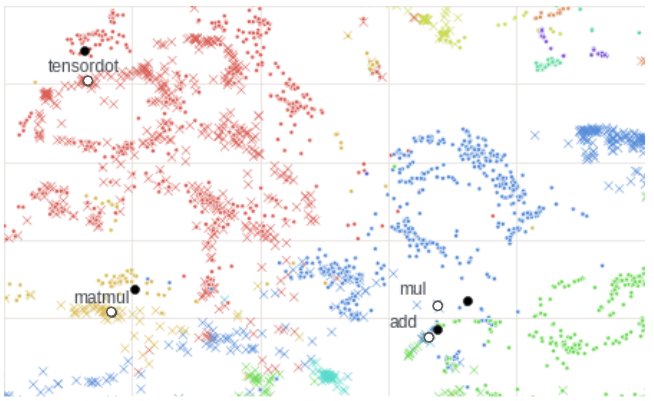}
    \caption{Proximity of $h_2$ and $h_2'$ pairs for some inputs (in white and black respectively), against a backdrop of $h_2$ (crosses) and $h_2'$ (dots).}
    \label{fig:h2h2p}
\end{figure}

The interesting property is that $h_2$ and $h_2'$ are close together in the representational space.  Figure~\ref{fig:h2h2p} shows a part of a tSNE plot of $h_2$ (crosses) and 
$h_2'$ (dots), as described above, for
inputs drawn from our benchmarks. 
As we expect, black and white markers in the figure show that $h2$ and $h2'$ for the same inputs are arranged close to
each other in the embedding space. In producing $h_2$ (or $h_2'$), the RNN unit did not care whether it was given $h_1$, the representation produced by the previous RNN unit, or directly given $f_1(inp_1)$.
In this manner, successive hidden states contain information analogous to the results of concrete computations: $f_1(inp_1) $,
$f_2 ( f_1( inp_1 ) ) $, and so on. (In the actual model, these functions need not be unary, as implied here.)

\paragraph*{Composition property formally.}
Refer to the notation introduced in Sec~\ref{sec:compositional}. If instead of $h_{i-1}$,
$\mathit{RNN}_i$ is  given (the embedding of) the concrete intermediate value computed by the computation so far, its observed behavior is the same in both cases. Formally,
{
\setlength{\abovedisplayskip}{0pt}
\setlength{\belowdisplayskip}{0pt}
\begin{equation}
\begin{split}
\mathbf{h}_i, d_i & = \mathit{RNN}_i ( \llbracket f_{i-1}(args_{i-1})\#\mathit{out} \rrbracket, \llbracket inp_{i}\#\mathit{out} \rrbracket) \\
& \approx \mathit{RNN}_i(\mathbf{h}_{i-1}, \llbracket inp_i \# out \rrbracket) \\
\end{split}
\end{equation}
}

Thus, the hidden vectors capture  an abstraction of intermediate values that would have arisen
in concrete computation $f_1(i_1), f_2(f_1(i_1))$, and so on.  
The composition takes place as each unit of RNN make a local decision based on incoming hidden
vector, which is set up to capture the result of the corresponding concrete computation so far.

\section{Limitations}
\label{sec:discussion}

Our results are promising, yet preliminary in many ways, and we have not established generality in
several dimensions.
First, we support a  small set of API functions and have carried out a limited evaluation.  
As the number increases, the training data size also increases, and training the model well becomes harder due to computational needs.  The robustness of training is a challenge in general.

Second, the model's ability to generalize to unseen sequences is crucially dependent on training over a broad diversity of API sequences.  This is challenging as we go to a larger number of API functions, because we cannot cover all permutations exhaustively.
However, as seen in Section~\ref{subsec:eval_scalability}, the model can still learn the semantics reasonably well if the training data covers the sequences in the test set.
Thus, the future works may benefit from creating a training dataset containing a distribution of sequences representing the real-world API usage patterns, through API usage mining~\cite{zhang2018code,nam2019marble}.

Third, we have explored the model's training and inference on relatively short tensors, with small data ranges, and have generally worked only with integer data.  In a real application, tensors can be out-of-distribution with respect to the model.   

Fourth, we have worked only with PyTorch.  We believe the work can be replicated easily to NumPy and Tensorflow API functions, because of their similar nature (acting over arrays of numbers.).  Farther out, we may need to invent additional techniques.  
For example, Pandas is designed to deal with records (or ``dataframes'') containing labeled axes (rows and columns); so it supports not only numeric manipulation, but also label-based slicing.  Here we may be able to borrow some insights from the Autopandas work~\cite{bavishi2019autopandas}.  Finding ways to deal with the above threats is in our future work.

\noindent

\section{Related Work}
\label{sec:related}
\balance
\textbf{Program Synthesis.} It has a rich literature, including example driven LISP code generation~\cite{shaw1975inferring, hardy1974automatic}, deductive synthesis from decades ago~\cite{manna1980deductive}, sketch completion using satisfiability~\cite{solar2006combinatorial}, bit-vector manipulations~\cite{jha2010oracle}, string processing~\cite{gulwani2011automating,parisotto2016neuro}, 
data processing~\cite{smith2016mapreduce, yaghmazadeh2018automated}, 
syntax transformations~\cite{rolim2017learning}, database queries~\cite{yaghmazadeh2017sqlizer}, 
data wrangling~\cite{feng2018program, feng2017component, le2014flashextract}, and the highly successful programming-by-examples system FlashFill~\cite{gulwani2011automating}.  
FlashFill uses enumerative synthesis, where a space of programs are explored in some order, until one that fulfils a requirement---typically one or more examples---is found.  
Transit~\cite{udupa2013transit} is another well-known work in enumerative synthesis, where the exploration is arranged in terms of finding sub-expressions in order of their \emph{costs}.  
Increasingly more costly expressions are attempted, using expressions 
previously computed.

\textbf{ML for Program Synthesis.} With  advances in ML,  researchers tried to adopt ML 
on top of the enumerative search for more efficient program synthesis~\cite{bavishi2019autopandas, shi2020tf, balog2016deepcoder, odena2020bustle, nye2019learning}. 
Our work is closely related to DeepCoder~\cite{balog2016deepcoder}, TF-Coder~\cite{shi2020tf} and BUSTLE~\cite{odena2020bustle}. 
Using prediction-guided enumerative synthesizer, they show the benefits of predicting API functions 
that are needed \textit{somewhere} given a synthesis instance. 
However, they all use a explicit 
featurization over these the input-output values, which is not easy to generalize to other programming languages.
Also, they only predict presence or absence of API functions, the prediction was only used to prioritize operations in the enumerative search, rather than directly predicting the API function(s) in sequence.
With the ML model guiding the search, BUSTLE takes an approach 
similar to ours, which gives feedback to search iteratively, whereas the models of DeepCoder or TF-Coder only give feedback in the beginning of the search.
However, BUSTLE and DeepCoder only support simple DSL tasks, which may not be generalized for real-world API-based synthesis.

\textbf{Neural Program Synthesis.}
Approaches like~\cite{devlin2017robustfill,bunel2018leveraging,parisotto2016neuro,balog2020neural,parisotto2016neuro,yin2017syntactic, ahmad2021unified} 
directly use neural networks 
for end-to-end synthesis~\cite{devlin2017robustfill,bunel2018leveraging,parisotto2016neuro,balog2020neural} 
to generate string transformation programs from examples. 
These works generally use encoder-decoder model.
In particular, the encoder embeds the input/output strings, and the decoder generates the program sequences conditioned on the input embedding. 
However, these approaches 
are mostly built and evaluated with simple DSL tasks, mostly with simple string transformation. 
In this work, we worked on the real-world tensor manipulation library PyTorch.
Although our evaluation does not cover the full range of PyTorch, we found several challenges in expanding these work into more complex programs, such as the scalability issue in training data generation and diversity of the API parameters especially in the tensor domain.

\textbf{Execution-guided Program Synthesis.}
Recent works have tried to exploit program execution to learn better representations for the neural program synthesis~\cite{shin2018improving, chen2021latent, bunel2018leveraging, ellis2019write, nye2020representing,chen2018execution}.
Some of the approaches~\cite{shin2018improving, bunel2018leveraging} use program interpreters to provide the actual intermediate execution results, and a more recent approach~\cite{chen2021latent} learns the latent representation to approximate the execution of partial programs using a separate ``Latent Executor''.
We also learn the representation of the execution of partial programs and demonstrate that (see Section~\ref{fig:equivalence2}). 
However, we capture the intermediate execution results as part of the main recurrent model, without needing to use the separate module to approximate or execute the program.

\noindent

\section{Conclusion}
\label{sec:conclusion}

In this paper, we proposed a new machine learning technique to speed up enumerative program synthesis.
Our idea is to use an ML model to predict the sequence of API function calls required to go from an input to the final desired output, in our case, both numeric vectors.
Our model is trained on randomly generated data.
It is able to predict API sequences for previously unseen inputs and outputs.  Moreover, it can 
predict API sequences that were not seen during training either.  The model does so by learning to compose API sequences, by learning how to keep track of values in the hidden states of an RNN.  We showed that our model can predict sequences of lengths 1 to 3 fairly well. In terms of effectiveness, we showed that our technique accelerates enumerative synthesis more effectively than related previous works DeepCoder~\cite{balog2016deepcoder} and TF-Coder~\cite{shi2020tf}.  




\balance
\clearpage
\newpage
\bibliographystyle{ACM-Reference-Format}
\bibliography{main, ray}
\clearpage

\onecolumn
\appendix
\newpage
\renewcommand{\thesection}{\Alph{section}}

\section{Supported operations of PyTorch}
Below is the list of 33 PyTorch operations. 16 operations used in the original dataset (described in Section~\ref{subsec:dataset}) are highlighted.

\begin{itemize}
    \item \textbf{add}
    \item \textbf{any}
    \item arange
    \item argmax
    \item \textbf{bincount}
    \item cdist
    \item div
    \item \textbf{eq}
    \item \textbf{expand}
    \item eye
    \item gather
    \item \textbf{gt}
    \item \textbf{lt}
    \item \textbf{masked\_select}
    \item \textbf{matmul}
    \item max
    \item minimum
    \item \textbf{mul}
    \item \textbf{ne}
    \item one\_hot
    \item repeat\_interleave
    \item reshape
    \item roll
    \item searchsorted
    \item square
    \item squeeze
    \item \textbf{stack}
    \item sum
    \item \textbf{tensordot}
    \item tile
    \item \textbf{transpose}
    \item \textbf{unsqueeze}
    \item \textbf{where}
\end{itemize}

\section{Stack Overflow Benchmarks}
As mentioned in Section~\ref{subsec:dataset}, we adapted the Stack Overflow benchmarks created for TF-Coder~\cite{shi2020tf}.
The examples were collected from Stack Overflow posts and the benchmarks were inspired by those posts.
However, the input/output values were replaced by the TF-Coder authors for licensing reasons.
The input/output values created by TF-Coder authors, and we updated some values to fit into our scope.
\newpage
\subsection{Input/output and Desired Code}
\begin{figure}[hbt]
    \centering
    \includegraphics[width=0.92\textwidth,page=1]{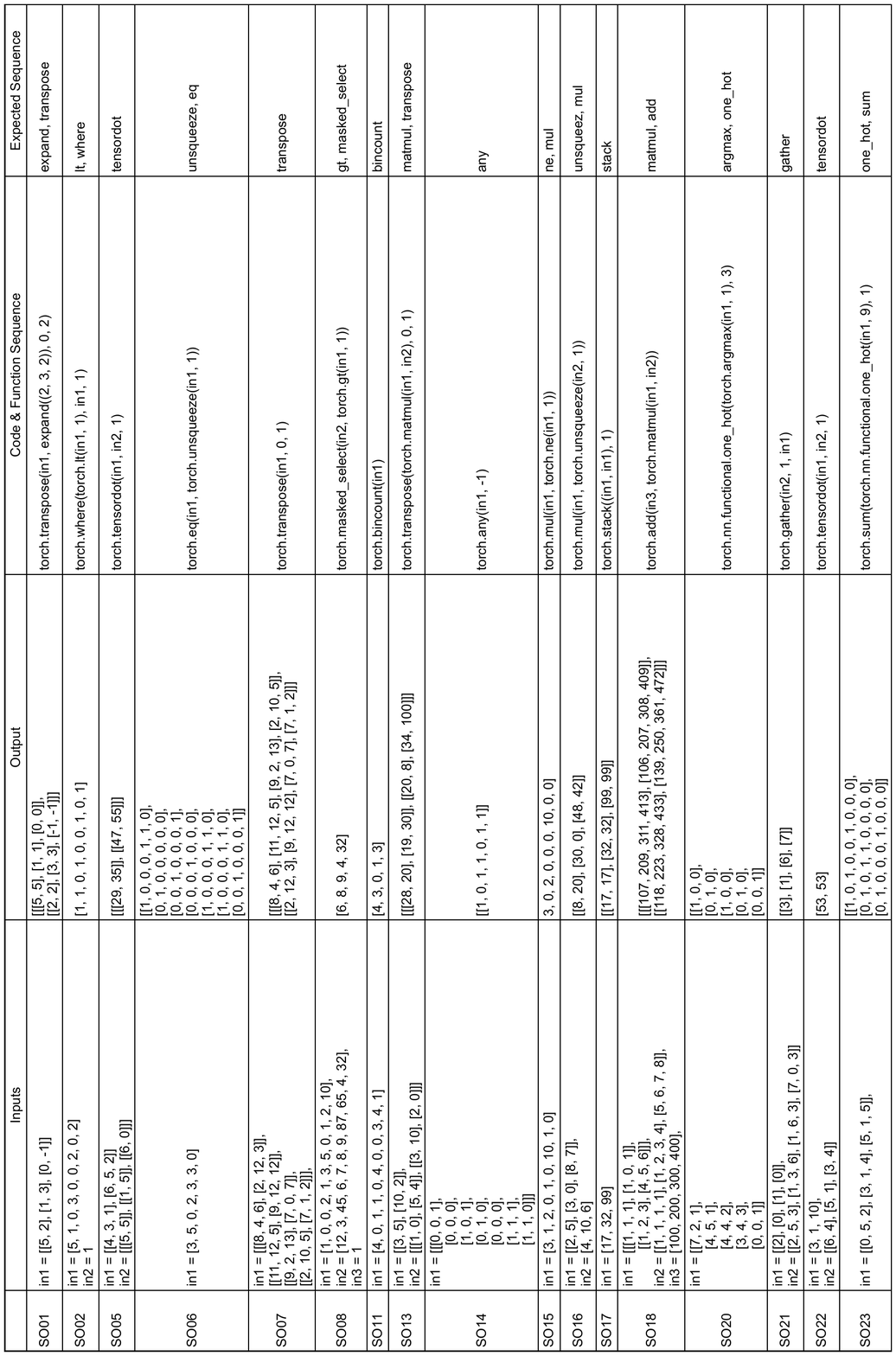}
\end{figure}
\begin{figure}[hbt]
    \centering
    \includegraphics[width=0.92\textwidth,page=2]{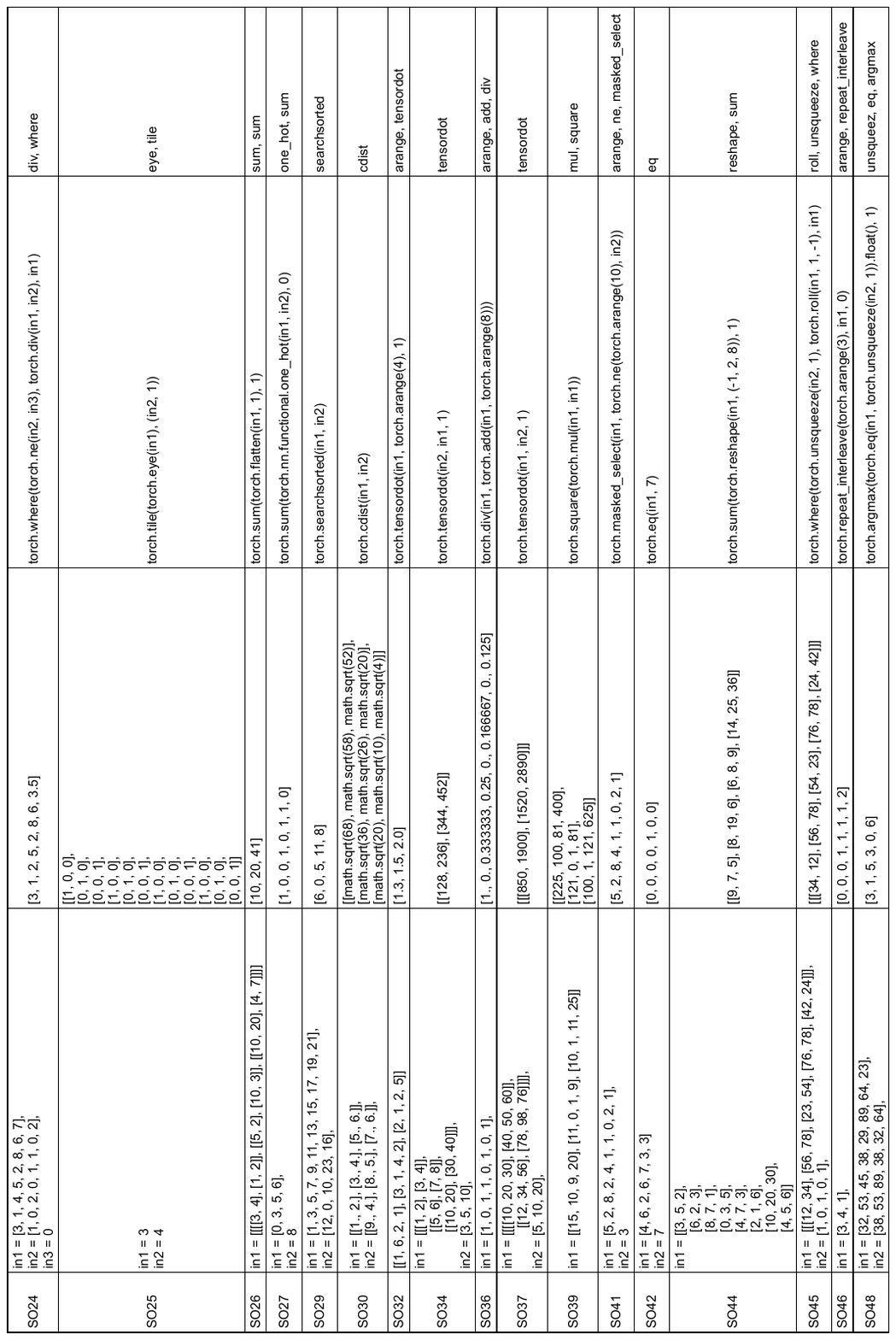}
\end{figure}
\begin{figure}[hbt]
    \centering
    \includegraphics[width=0.92\textwidth,page=3]{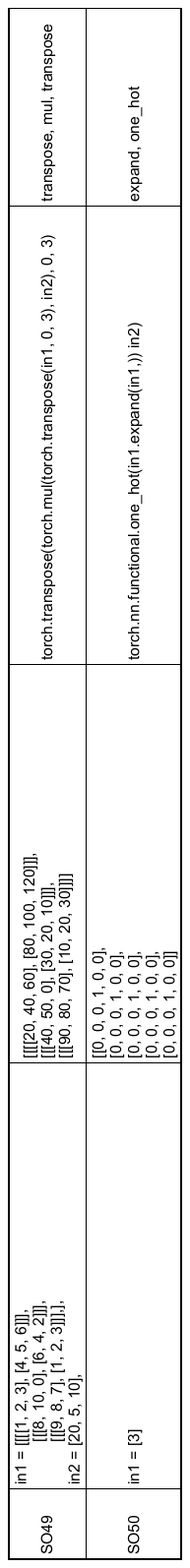}
\end{figure}

\newpage
\clearpage
\subsection{Links to Original StackOverflow Posts}
\begin{itemize}
\item \url{https://stackoverflow.com/questions/40441503/tensorflow-tensor-reshape}
\item \url{https://stackoverflow.com/questions/46408839/tensorflow-trim-values-in-tensor}
\item \url{https://stackoverflow.com/questions/43067338/tensor-multiplication-in-tensorflow}
\item \url{https://stackoverflow.com/questions/47816231/create-binary-tensor-from-vector-in-tensorflow}
\item \url{https://stackoverflow.com/questions/38212205/swap-tensor-axes-in-tensorflow}
\item \url{https://stackoverflow.com/questions/33769041/tensorflow-indexing-with-boolean-tensor}
\item \url{https://stackoverflow.com/questions/45194672/how-to-count-elements-in-tensorflow-tensor}
\item \url{https://stackoverflow.com/questions/50777704/n-d-tensor-matrix-multiplication-with-tensorflow}
\item \url{https://stackoverflow.com/questions/35657003/aggregate-each-element-of-tensor-in-tensorflow}
\item \url{https://stackoverflow.com/questions/39045797/conditional-assignment-of-tensor-values-in-tensorflow}
\item \url{https://stackoverflow.com/questions/46240646/tensor-multiply-along-axis-in-tensorflow}
\item \url{https://stackoverflow.com/questions/51761353/about-tensor-of-tensorflow}
\item \url{https://stackoverflow.com/questions/38222126/tensorflow-efficient-way-for-tensor-multiplication}
\item \url{https://stackoverflow.com/questions/44834739/argmax-on-a-tensor-and-ceiling-in-tensorflow}
\item \url{https://stackoverflow.com/questions/51690095/how-to-gather-element-with-index-in-tensorflow}
\item \url{https://stackoverflow.com/questions/43284897/how-can-i-multiply-a-vector-and-a-matrix-in-tensorflow-without-reshaping}
\item \url{https://stackoverflow.com/questions/53414433/tensorflow-tensor-binarization}
\item \url{https://stackoverflow.com/questions/53643339/tensorflow-overriding-tf-divide-to-return-the-numerator-when-dividing-by-0}
\item \url{https://stackoverflow.com/questions/53602691/duplicate-a-tensor-n-times}
\item \url{https://stackoverflow.com/questions/54294780/how-to-perform-reduce-op-on-multiple-dimensions-at-once}
\item \url{https://stackoverflow.com/questions/54225704/how-do-i-get-a-tensor-representing-the-on-positions-in-the-original-tensor}
\item \url{https://stackoverflow.com/questions/54155085/bucketing-continous-value-tensors-in-tensorflow}
\item \url{https://stackoverflow.com/questions/54147780/tensorflow-how-to-calculate-the-euclidean-distance-between-two-tensor}
\item \url{https://stackoverflow.com/questions/48659449/how-to-compute-the-weighted-sum-of-a-tensor-in-tensorflow}
\item \url{https://stackoverflow.com/questions/49532371/compute-a-linear-combination-of-tensors-in-tensorflow}
\item \url{https://stackoverflow.com/questions/43306788/divide-elements-of-1-d-tensor-by-the-corrispondent-index}
\item \url{https://stackoverflow.com/questions/49206051/multiply-4-d-tensor-with-1-d-tensor}
\item \url{https://stackoverflow.com/questions/37912161/how-can-i-compute-element-wise-conditionals-on-batches-in-tensorflow}
\item \url{https://stackoverflow.com/questions/54499051/elegant-way-to-access-python-list-and-tensor-in-tensorflow}
\item \url{https://stackoverflow.com/questions/54493814/binary-vector-of-max}
\item \url{https://stackoverflow.com/questions/54402389/sum-the-columns-for-each-two-consecutive-rows-of-a-tensor-of-3-dimensions}
\item \url{https://stackoverflow.com/questions/54337925/reverse-order-of-some-elements-in-tensorflow}
\item \url{https://stackoverflow.com/questions/58652161/how-to-convert-2-3-4-to-0-0-1-1-1-2-2-2-2-to-utilize-tf-math-segment-sum}
\item \url{https://stackoverflow.com/questions/58481332/getting-the-indices-of-several-elements-in-a-tensorflow-at-once}
\item \url{https://stackoverflow.com/questions/58466562/given-a-batch-of-n-images-how-to-scalar-multiply-each-image-by-a-different-scal}
\item \url{https://stackoverflow.com/questions/58537495/tensorflow-initialize-a-sparse-tensor-with-only-one-line-column-not-zero}    
\end{itemize}

\newpage


\newpage
\section{Implementation of ML Models}
\label{sec:implementation}

We implement the model in Python using the PyTorch. 
We describe some implementation details here, but the code will be shared after anonymous period is over.

\textbf{Encoding.} 
Encoding of a tensor requires three encodings, separated by a separator: value, size, and type. The max sizes of the encodings are 150, 5, 3, respectively. 
Our approach supports up to 3 input tensors and one output tensor, and each tensor is separated by a separator, which makes the the size of input encoding to be
640 ($4*(150+1+5+1+2+1)$).



\textbf{Compositional Model.} For the embedding, we use the feed forward network, that is identical to the classification model, which is trained jointly with a bi-RNN model.
The embedded input-output pair is passed to bi-RNN, having 1 hidden layer.
To evaluate the model, we use a beam size of 3. 
We use the compositional model in two modes: ``full sequence'' mode, which returns the predicted API function sequence, and a ``first-of-sequence'' mode that returns only the first API function from the predicted sequence.



\textbf{Multi-label Classification Model.}
For the  weighted enumerative search with prioritization (Figure~\ref{fig:overview} (b)), we trained a simple multi-label classification model following DeepCoder, instead of TF-Coder which uses  manually defined features (e.g., whether a value is a primitive) which we found less generalizable.
We used the same model architecture with the classification model, but changed the last activation function into \textit{sigmoid} for the multi-label classification. 
We trained this model with input-output of \textit{sequences} of API functions. 

\textbf{Evaluation Metrics.} We evaluate our models on the accuracy of their predictions.
For the model accuracy with synthetic data, we check whether the model correctly predict the ground-truth APIs, and for the Stack Overflow benchmarks evaluation, we measure the rank of the correct prediction and extract top-1, top-3 and top-10 accuracy metrics.


\section{Algorithms}

\begin{algorithm}[h!]
\caption{Weighted Enumerative Synthesis}
\label{alg:iterative}
\begin{algorithmic}[1]
\Require{A task specification input/output, ($I, O$)} 
\Ensure{A program $P$ such that $P(I) = O$}
\Statex
\State{$B \gets \lbrace I, 0, -1, 1, ... \rbrace$}  \Comment{Base values}
\State {$E \gets B$} \Comment{Pool of values}
\State{$Ops \gets \mathit{AssignOpCost}(I, O)$}
\ForAll {$v \in C$}
    \State {$v.cost \gets \mathit{AssignValCost}(v)$}
    \EndFor
\For {$C = 1 \xrightarrow{} max\_cost$}   \Comment{Cost Budget}
    \ForAll{$op \in Ops$}
        \State{$c \gets op.cost$}
        \State{$n \gets op.arity$}
        \LineComment{Partition cost budgets into $n$ arguments}
        \ForAll{$[c_1,...,c_n] \in partition(C-c, n)$}
            \For{$i=1,...,n$}
            \LineComment{Collect values satisfying i-th arg cost budget}
                \State{$A_i \gets \lbrace e \in E | e.cost = c_i \rbrace $}
            \EndFor
            \ForAll{$args \in \Pi_i A_i$}
                \State {$V \gets Execute(op, args)$}    \Comment{Run $op$ w/ $args$}
                \If {$V = O$}
                    \Return {$\mathit{V.expr}$}
                \EndIf
                \If {$V \notin E$}
                    \State{$V.cost \gets C$}
                    \State{$E \gets E \cup \{V\}$}
                \EndIf
            \EndFor
        \EndFor
    \EndFor
\EndFor
\State \Return {"Fail: reached maximum cost"}
\end{algorithmic}
\end{algorithm}

\begin{algorithm}[h!]
\caption{AssignOpCost}
\label{alg:assignopweight}
\begin{algorithmic}[1]
\Require{A task specification input/output, ($I, O$)} 
\Ensure{List of operations with costs $Ops$}
\Statex
\ForAll{$op \in Ops$}
    \State{$op \gets preset\_cost$}
\EndFor
\If {doModelPrioritization}
    \State{$candidate\_ops \gets \mathit{MultiClassificationModel}(I, O)$}
    \ForAll{$op \in \mathit{candidate\_ops}$}
        \State{$op.cost \gets op.cost * reweight\_multiplier$}
    \EndFor
\EndIf\\
\Return{$Ops$}
\end{algorithmic}
\end{algorithm}



\begin{algorithm}[h!]
\caption{Compositional Model - Full-Sequence}
\label{alg:full}
\begin{algorithmic}[1]
\Require{A task specification input/output, ($I, O$)} 
\Ensure{A program $P$ such that $P(I) = O$}
\Statex
\State{$B \gets \lbrace I, 0, -1, 1, ... \rbrace$}

\State {$op\_seq \gets \mathit{CompositionalModel}(I, O)$}
\State {$n \gets \sum op_i.arity$}
\State{$args\_list \gets \Pi_n B$}
\ForAll{$args \in args\_list$}
    \State {$V \gets Execute(op\_seq, args)$}
    \If {$V = O$}
        \Return {$\mathit{V.expr}$}
\EndIf
\EndFor
\State \Return {"Fail" or \textbf{start} "Enumerative Search"}
\end{algorithmic}
\end{algorithm}

\begin{algorithm}[h!]
\caption{Compositional Model - First-Of-Sequence}
\label{alg:fos}
\begin{algorithmic}[1]
\Require{A task specification input/output, ($I, O$)} 
\Ensure{A program $P$ such that $P(I) = O$}
\Statex
\State{$B \gets \lbrace I, 0, -1, 1, ... \rbrace$}
\For {$i = 0 \xrightarrow{} k$} \Comment{Sequence of $k$ operations}
    \State {$op_i \gets CompositionalModel(V_{i-1}, I_i, O)$}
    \State {$n \gets op_i.arity$}
    \State{$args\_list \gets \Pi_n B$}
    \ForAll{$args \in args\_list$}
        \State {$V_i \gets Execute(op_i, args)$}
        \If {$V_i = O$}
            \Return {$\mathit{V.expr}$}
        \EndIf
    \EndFor
\EndFor
\State \Return {"Fail" or \textbf{start} "Enumerative Search"}
\end{algorithmic}
\end{algorithm}

\end{document}